*Research Article*

# Mixing Energy Models in Genetic Algorithms for On-Lattice Protein Structure Prediction

**Mahmood A. Rashid,[1,2] M. A. Hakim Newton,[1] Md. Tamjidul Hoque,[3] and Abdul Sattar[1,2]**

[1] *Institute for Integrated & Intelligent Systems, Science 2 (N34) 1.45, 170 Kessels Road, Nathan, QLD 4111, Australia*
[2] *Queensland Research Lab, National ICT Australia, Level 8, Y Block, 2 George Street, Brisbane, QLD 4000, Australia*
[3] *Computer Science, 2000 Lakeshore drive, Math 308, New Orleans, LA 70148, USA*

Correspondence should be addressed to Mahmood A. Rashid; mahmood.rashid@gmail.com

Protein structure prediction (PSP) is computationally a very challenging problem. The challenge largely comes from the fact that the energy function that needs to be minimised in order to obtain the native structure of a given protein is not clearly known. A high resolution $20 \times 20$ energy model could better capture the behaviour of the actual energy function than a low resolution energy model such as hydrophobic polar. However, the fine grained details of the high resolution interaction energy matrix are often not very informative for guiding the search. In contrast, a low resolution energy model could effectively bias the search towards certain promising directions. In this paper, we develop a genetic algorithm that mainly uses a high resolution energy model for protein structure evaluation but uses a low resolution HP energy model in focussing the search towards exploring structures that have hydrophobic cores. We experimentally show that this mixing of energy models leads to significant lower energy structures compared to the state-of-the-art results.

## 1. Introduction

Proteins are essentially sequences of amino acids. They adopt specific folded three-dimensional structures to perform specific tasks. However, misfolded proteins cause many critical diseases such as Alzheimer's disease, Parkinson's disease, and cancer [1, 2]. Protein structures are important in drug design and biotechnology.

Protein structure prediction (PSP) is computationally a very hard problem [3]. Given a protein's amino acid sequence, the problem is to find a three-dimensional structure of the protein such that the total interaction energy amongst the amino acids in the sequence is minimised. The protein folding process that leads to such structures involves very complex molecular dynamics [4] and unknown energy factors. To deal with the complexity of PSP in a hierarchical way, researchers have used discretised lattice-based structures and simplified energy models [5–7].

There are a large number of existing search algorithms that attempt to solve the PSP problem by exploring feasible structures called *conformations*. For the low resolution hydrophobic-polar (HP) energy model, a memory based local search algorithm [8, 9], a population-based genetic algorithm [10], and a hydrophobic core directed local search method [11] reportedly produced the state-of-the-art results on the face-centred-cubic (FCC) lattice. For the high resolution Berrera $20 \times 20$ energy matrix (henceforth referred to as BM energy model) [12–14] produces the state-of-the-art results. Nevertheless, the challenges in PSP largely remain in the fact that the energy function that needs to be minimised in order to obtain the native structure of a given protein is not clearly known. A high resolution $20 \times 20$ energy model (such as BM) could better capture the behaviour of the actual energy function than a low resolution energy model (such as HP). However, the fine grained details of the high resolution interaction energy matrix are often not very informative for guiding the search. Pairwise contributions that have large magnitudes could be overshadowed by the accumulation of pair-wise contributions having small magnitudes or opposite signs. In contrast, a low resolution energy model could effectively bias the search towards certain promising directions particularly emphasising the pair-wise contributions with large magnitudes.

In this paper, we present a genetic algorithm that mainly uses a high resolution energy model for protein structure evaluation but uses a low resolution HP energy model in focussing the search towards exploring structures that have hydrophobic cores. Protein structures have hydrophobic cores that hide the hydrophobic amino acids from water and expose the polar amino acids to the surface to be in contact with the surrounding water molecules [15]. We apply a macromutation operator that considers the HP energy model and attempts to build hydrophobic cores. We experimentally show that our way of mixing these two energy functions leads to significant lower energy structures compared to the state-of-the-art results.

The rest of the paper is organised as follows: Section 2 reviews background knowledge; Section 3 discusses related work on PSP; Section 4 describes our methods in detail; Section 5 presents the experimental results and analyses; and finally, Section 6 draws the conclusions and outlines the future research.

## 2. Background

There are three computational approaches for protein structure prediction. They are *homology modeling*, *protein threading*, and *ab initio* approach. The prediction quality of *homology modeling* depends on the sequential similarity with proteins that have previously known structures. On the other hand, protein threading, also known as fold recognition, is based on the structural similarity with the previously known fold families. Our work is based on the *ab initio* approach that depends only on the amino acid sequence of the target protein. Levinthal's paradox [16] and Anfinsen's hypothesis [17] are the basis of *ab initio* method for protein structure prediction. The idea was originated in 1970 when it was demonstrated that all information needed to fold a protein, resides in its amino acid sequence; that is, given the solvent properties, amino acids in the sequence determine the unique fold of a protein.

To explore an astronomically large search space and to evaluate the conformations using a real energy model are a big challenge for existing search algorithms in PSP. Therefore, simplified models have significant importance in understanding the protein folding process.

### 2.1. Simplified Model.
In our simplified model, we use 3D FCC lattice points to map the amino acids of a protein sequence. In the mapping, each amino acid of the sequence occupies a point on the lattice to form a continuous chain of self-avoiding walk. We use the BM and HP models together within a population-based genetic algorithm (GA) for protein structure prediction. The FCC lattice, the HP and BM energy models, and the GA are described below.

#### 2.1.1. 3D FCC Lattice.
The FCC lattice has the highest packing density compared to the other existing lattices [18].

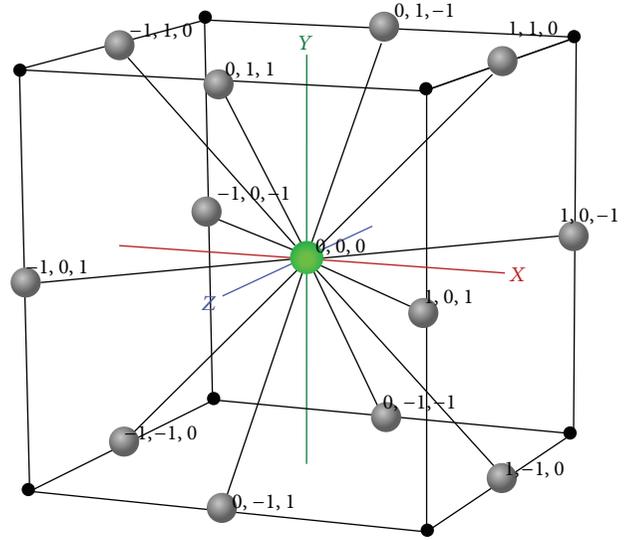

Figure 1: A unit of 3D FCC lattice with 12 basis vectors on the Cartesian space.

In FCC, each lattice point (e.g., the origin in Figure 1) has 12 neighbours with 12 *basis vectors* as follows:

$$v_1 = (1, 1, 0), \quad v_2 = (-1, -1, 0), \quad v_3 = (-1, 1, 0),$$
$$v_4 = (1, -1, 0), \quad v_5 = (0, 1, 1), \quad v_6 = (0, 1, -1),$$
$$v_7 = (1, 0, 1), \quad v_8 = (1, 0, -1), \quad v_9 = (0, -1, 1),$$
$$v_{10} = (-1, 0, 1), \quad v_{11} = (0, -1, -1), \quad v_{12} = (-1, 0, -1).$$
(1)

In simplified PSP, conformations are mapped on the lattice by a sequence of basis vectors or by the *relative vectors* that are relative to the previous basis vectors in the sequence.

#### 2.1.2. HP Energy Model.
Twenty different amino acids are the primary constituents of proteins. Based on the hydrophobic property, these 20 amino acids are broadly divided into two categories: (a) hydrophobic amino acids (*Gly*, *Ala*, *Pro*, *Val*, *Leu*, *Ile*, *Met*, *Phe*, *Tyr*, and *Trp*) denoted by H and (b) hydrophilic or polar amino acids (*Ser*, *Thr*, *Cys*, *Asn*, *Gln*, *Lys*, *His*, *Arg*, *Asp*, and *Glu*) denoted by P. In the HP model [19], when two nonconsecutive hydrophobic amino acids become topologically neighbours, they contribute a certain amount of negative energy, which for simplicity is considered as −1.

The total energy $E_{HP}$ (as shown in (2)) of a conformation based on the HP model becomes the sum of the contributions over all pairs of non-consecutive hydrophobic amino acids:

$$E_{HP} = \sum_{i < j-1} c_{ij} \cdot e_{ij},$$
(2)

where $c_{ij} = 1$ if amino acids at positions $i$ and $j$ in the sequence are non-consecutive neighbours on the lattice, otherwise 0, and $e_{ij} = -1$ if $i$th and $j$th amino acids are both hydrophobic, otherwise 0.

*2.1.3. BM Energy Model.* By analysing crystallised protein structures, Miyazawa and Jernigan [21] in 1985 statistically deduced a 20 × 20 energy matrix that considers residue contact propensities between the amino acids. By calculating empirical contact energies on the basis of the information available from a huge number of selected protein structures and following the quasi-chemical approximation, Berrera et al. [20] in 2003 deduced another 20 × 20 energy matrix. In this work, we use the latter model and denote it by BM energy model. Table 1 shows the BM energy model with amino acid names at the leftmost column and the bottommost row and the interaction energy values in the cells. The amino acid names that have boldface are hydrophobic. We draw lines in Table 1 to show groupings based on H-H, H-P, and P-P interactions. In the context of this work, it is worth noting that most energy contributions that have large magnitudes are from H-H interactions followed by those from H-P interactions.

The total energy $E_{BM}$ (shown in (3)) of a conformation based on the BM energy model is the sum of the contributions over all pairs of non-consecutive amino acids that are one unit lattice distance apart:

$$E_{BM} = \sum_{i<j-1} c_{ij} \cdot e_{ij}, \quad (3)$$

where $c_{ij} = 1$ if amino acids at positions $i$ and $j$ in the sequence are non-consecutive neighbours on the lattice, otherwise 0, and $e_{ij}$ is the empirical energy value between the $i$th and $j$th amino acid pair specified in the matrix for the BM model.

*2.2. Genetic Algorithms.* GAs are a family of population-based search algorithms for optimisation problems. GAs maintain a set of solutions known as population. In each *generation*, it generates a new population from the current population using a given set of genetic operators known as *crossover* and *mutation*. It then replaces the inferior solutions by superior newly generated solutions to get a better current population. The generic pseudocode of GA is presented in Algorithm 1. A typical crossover operator randomly splits two solutions at a randomly selected crossover point and exchanges parts between them (Figure 2(a)) and a typical mutation operator alters a solution at a random point (Figure 2(b)). In the case of PSP, conformations are regarded as solutions of a GA population.

*2.2.1. Crossover Operators.* The crossover operators are applied on two selected parent conformations to exchange their parts to generate child conformations. In a *single-point crossover*, both parents are split at a single point (Figure 2(a)) while in a *multipoint crossover* they are split at more than one point. Nevertheless, the crossover operations succeed if they produce conformations that satisfy the self-avoiding walk constraint. In lattice-based protein representation, a self-avoiding walk constraint ensures no revisitation of any lattice point during the sequence mapping.

*2.2.2. Mutation Operators.* The mutation operators (Figure 2(b)) are applied on a single conformation. The operators can perform single-point change or multi-point changes. The mutation operations succeed if the resultant conformation remains a self-avoiding walk on the lattice.

## 3. Related Work

We explored the literature on protein structure prediction based on both HP and 20 × 20 energy models.

*3.1. HP Energy Based Approaches.* Different types of meta-heuristic have been used in solving the simplified PSP problem. These include Monte Carlo simulation [22], simulated annealing [23], genetic algorithms (GA) [24, 25], tabu search with GA [26], tabu search with hill climbing [27], ant colony optimisation [28], particle swarm optimisation [29, 30], immune algorithms [31], tabu-based stochastic local search [8, 32], and constraint programming [33, 34].

Cebrián et al. [32] used tabu-based local search, and Shatabda et al. [8] used memory-based local search with tabu heuristic and achieved the state-of-the-art results. However, Dotu et al. [34] used constraint programming and found promising results. A constraint programming based exact and complete algorithm for structure prediction is implemented in the CPSP tools by Mann et al. [33]. These algorithms can find the optimal solution if the target protein has a matching hydrophobic-core stored in the CPSP database.

Besides local search, Unger and Moult [24] applied genetic algorithms to PSP and found their method to be more promising than the Monte Carlo based methods [22]. The GA has been used by Hoque et al. [35] for cubic and 3D HCP lattices. They also introduced a twin-removal operator [36] to remove duplicates from the GA population.

However, using 3D FCC lattice points, the recent state-of-the-art results for HP energy models have been achieved by genetic algorithms [10, 37], local search approaches [8, 11], a local search embedded GA [38], and a multipoint parallel local search approach [39].

*3.2. Empirical 20 × 20 Matrix Energy Based Approaches.* A constraint programming technique was used in [40] by Dal Palù et al. to predict tertiary structures of real proteins using secondary structure information. They also used constraint programming with different heuristics in [41] and a constraint solver named COLA [42] that is highly optimised for protein structure prediction. In another work [43], a fragment assembly method was utilised with empirical energy potentials to optimise protein structures. Among other successful approaches, a population based local search [44] and a population based genetic algorithm [14] were used with empirical energy functions.

In a hybrid approach, Ullah and Steinhöfel [12] applied a constraint programming based large neighbourhood search technique on top of the of COLA solver. The hybrid approach produced the state-of-the-art results for several small sized (less than 75 amino acids) benchmark proteins. In another

Table 1: The 20 × 20 BM energy model by Berrera et al. [20].

| | Cys | Met | Phe | Ile | Leu | Val | Trp | Tyr | Ala | Gly | Thr | Ser | Gln | Asn | Glu | Asp | His | Arg | Lys | Pro |
|---|---|---|---|---|---|---|---|---|---|---|---|---|---|---|---|---|---|---|---|---|
| **Cys** | −3.477 | | | | | | | | | | | | | | | | | | | |
| **Met** | −2.24 | −1.901 | | | | | | | | | | | | | | | | | | |
| **Phe** | −2.424 | −2.304 | −2.467 | | | | | | | | | | | | | | | | | |
| **Ile** | −2.41 | −2.286 | −2.53 | −2.691 | | | | | | | | | | | | | | | | |
| **Leu** | −2.343 | −2.208 | −2.491 | −2.647 | −2.501 | | | | | | | | | | | | | | | |
| **Val** | −2.258 | −2.079 | −2.391 | −2.568 | −2.447 | −2.385 | | | | | | | | | | | | | | |
| **Trp** | −2.08 | −2.09 | −2.286 | −2.303 | −2.222 | −2.097 | −1.867 | | | | | | | | | | | | | |
| **Tyr** | −1.892 | −1.834 | −1.963 | −1.998 | −1.919 | −1.79 | −1.834 | −1.335 | | | | | | | | | | | | |
| **Ala** | −1.7 | −1.517 | −1.75 | −1.872 | −1.728 | −1.731 | −1.565 | −1.318 | −1.119 | | | | | | | | | | | |
| **Gly** | −1.101 | −0.897 | −1.034 | −0.885 | −0.767 | −0.756 | −1.142 | −0.818 | −0.29 | 0.219 | | | | | | | | | | |
| **Thr** | −1.243 | −0.999 | −1.237 | −1.36 | −1.202 | −1.24 | −1.077 | −0.892 | −0.717 | −0.311 | −0.617 | | | | | | | | | |
| **Ser** | −1.306 | −0.893 | −1.178 | −1.037 | −0.959 | −0.933 | −1.145 | −0.859 | −0.607 | −0.261 | −0.548 | −0.519 | | | | | | | | |
| **Gln** | −0.835 | −0.72 | −0.807 | −0.778 | −0.729 | −0.642 | −0.997 | −0.687 | −0.323 | 0.033 | −0.342 | −0.26 | 0.054 | | | | | | | |
| **Asn** | −0.788 | −0.658 | −0.79 | −0.669 | −0.524 | −0.673 | −0.884 | −0.67 | −0.371 | −0.23 | −0.463 | −0.423 | −0.253 | −0.367 | | | | | | |
| **Glu** | −0.179 | −0.209 | −0.419 | −0.439 | −0.366 | −0.335 | −0.624 | −0.453 | −0.039 | 0.443 | −0.192 | −0.161 | 0.179 | 0.16 | 0.933 | | | | | |
| **Asp** | −0.616 | −0.409 | −0.482 | −0.402 | −0.291 | −0.298 | −0.613 | −0.631 | −0.235 | −0.097 | −0.382 | −0.521 | 0.022 | −0.344 | 0.634 | 0.179 | | | | |
| **His** | −1.499 | −1.252 | −1.33 | −1.234 | −1.176 | −1.118 | −1.383 | −1.222 | −0.646 | −0.325 | −0.72 | −0.639 | −0.29 | −0.455 | −0.324 | −0.664 | −1.078 | | | |
| **Arg** | −0.771 | −0.611 | −0.805 | −0.854 | −0.758 | −0.664 | −0.912 | −0.745 | −0.327 | −0.05 | −0.247 | −0.264 | −0.042 | −0.114 | −0.374 | −0.584 | −0.307 | 0.2 | | |
| **Lys** | −0.112 | −0.146 | −0.27 | −0.253 | −0.222 | −0.2 | −0.391 | −0.349 | 0.196 | 0.589 | 0.155 | 0.223 | 0.334 | 0.271 | −0.057 | −0.176 | 0.388 | 0.815 | 1.339 | |
| **Pro** | −1.196 | −0.788 | −1.076 | −0.991 | −0.771 | −0.886 | −1.278 | −1.067 | −0.374 | −0.042 | −0.222 | −0.199 | −0.035 | −0.018 | 0.257 | 0.189 | −0.346 | −0.023 | 0.661 | 0.129 |

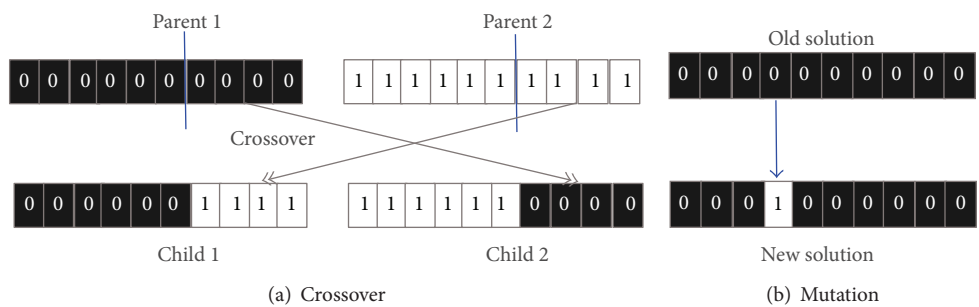

(1) initialise population;
(2) evaluate population;
(3) **while** (*!stopCondition*) **do**
(4)    select the best-fit individuals for reproduction;
(5)    breed new individuals through crossover and mutation operations;
(6)    evaluate the individual fitness of new individuals;
(7)    replace least-fit population with new individuals;

Algorithm 1: Generic pseudocode of a genetic algorithm.

(a) Crossover

(b) Mutation

Figure 2: Typical (a) crossover and (b) mutation operators.

work, Ullah et al. [45] proposed a two-stage optimisation approach combining constraint programming and local search. The first stage of the approach produced compact optimal structures by using the CPSP tools based on the HP model. In the second stage, those compact structures were used as the input of a simulated annealing based local search that is guided by the BM energy model.

In a recent work [13], Shatabda et al. presented a mixed heuristic local search algorithm for PSP and produced the state-of-the-art results using BM energy model on 3D FCC lattice. The mixed heuristic local search in each iteration randomly selects a heuristic from a given number of heuristics designed by the authors. The selected heuristics are then used in evaluating the generated neighbouring solutions of the current solution. Although the heuristics themselves are weaker than the BM energy, their collective use in the random mixing fashion produces results better than that produced by using the BM energy itself.

In this paper, we use the hydrophobic-polar categorisation of the HP model within a hydrophobic-core directed macromutation operator; however, most of the time the search is guided by the BM energy model. The macromutation is applied just like another mutation operator and utilises the distance of the hydrophobic amino acids from the hydrophobic core centre. The hydrophobic core centre is the centroid of only the hydrophobic amino acids. In effect the macro-mutation operator helps explore the structures that would have more interactions between H-H amino acid pairs; these pairs have energy contributions with large magnitudes. In the Results and Analyses section, we compare our experimental results with the results produced by [12, 13].

## 4. Methods and Implementations

In our genetic algorithm based *ab initio* protein structure prediction method, we use the BM energy model along with the HP energy model. We use face-centred-cubic lattice points for protein structure representation. In this section, we describe the implementation details of our GA framework.

*4.1. Genetic Algorithm Framework.* The *pseudocode* of our genetic algorithm framework is presented in Algorithm 2. It uses an exhaustive generation approach to diversify the search, a hydrophobic core directed macro-mutation operator to intensify the search, and a random-walk algorithm to recover from stagnation. The algorithm initialises (Line 6) the current population. At each generation, it selects a genetic operator based on a given probability distribution to use through the generation (Line 8). In fact, we select the operators randomly giving equal opportunities to all operators. The selected operator is used in an exhaustive way (Lines 10-11 or Lines 13–15) to obtain all conformations in the new population. We ensure that no duplicate conformation is added to the new population. The add( ) method either at Line 11 or at Line 15 also takes care of avoiding the duplicates while adding solutions to the new population list. For a given number of generations, if the best conformation in the new population is not better than the best in the current population, our algorithm then resorts to a random-walk procedure (Line 17) to diversify the new population. Nevertheless, after each generation, the new population becomes the current population (Line 18), and the search continues. Finally, the best conformation found so far is returned (Line 19).

Along with the BM energy mode, we use HP energy model to guide a macro-mutation. The macro-mutation operator is used as other mutation operators (Figures 4(b)–4(e)) in our GA. The exhaustive generation and macro-mutation are described below.

*4.1.1. Exhaustive Generation.* Unlike a traditional genetic algorithm, in our GA, the randomness is reduced significantly by an exhaustive generation approach. For mutation

```
(1) op: Operators, c, c': Conformations
(2) opR: Operator selection probabilities
(3) curP, newP: Current and new populations
(4) rwT: Number of non-improving generations before random walk
(5) //======================
(6) initPopulation(curP)
(7) foreach gereration until timeout do
(8)     selectOperator(op, opR)
(9)     if mutation(op) then
(10)        foreach c ∈ curP do
(11)            newP.add(mutConf(c))
(12)    else
(13)        while ¬ full(newP) do
(14)            c, c' ⟵ randomConfs(curP)
(15)            newP.add(crossConf(c, c'))
(16)    if ¬ improved(newP, rwT) then
(17)        randomWalk(newP)
(18)    curP ⟵ newP
(19) return bestConformation(curP)
```

ALGORITHM 2: The *pseudocode* of GA framework: geneticAlgorithm( ).

```
(1) mutantsadd(conf)
(2) foreach 1 ≤ pos ≤ seqLength do
(3)     c ⟵ applyOperator(conf pos)
(4)     mutantsadd(c)
(5) return bestConformation(mutants)
```

ALGORITHM 3: The *pseudocode* of exhaustive mutation: mutConf(conf).

operators, our GA adds one resultant conformation for *each* conformation in the current population to the new population. Operators are applied to all possible points (Algorithm 3) exhaustively until finding a better solution than the parent. If no better solution is found, the parent survives through the next generation. On the other hand, for crossover operators, two resultant conformations are added to the new population from two randomly selected parent conformations. Crossover operators generate child conformations by applying the crossover operator in all possible points (Algorithm 4) on two randomly selected parents. The best two conformations from the parents and the children are then become the resultant conformations for the next generation.

*4.1.2. Macromutation Operator.* Protein structures have hydrophobic cores that hide the hydrophobic amino acids from water and expose the polar amino acids to the surface to be in contact with the surrounding water molecules [15]. H-core formation is an important objective of HP based PSP. Macromutation operator is a composite operator (Figure 3) that uses a series of diagonal moves (Figure 4(c)) on a given conformation to build the H-core around the hydrophobic core center (HCC). The HCC is calculated by finding arithmetic means of $x$, $y$, and $z$ coordinates of all hydrophobic amino acids. The diagonal moves are applied repeatedly either at each P- or at each H-type amino acid position. Whether to apply the diagonal move on P- or H-type amino acids is determined by using a *Bernoulli* distribution (Algorithm 5 Line 2) with probability $p$ (intuitively we use $p = 20\%$ for P-type amino acids). For a P-type amino acid, the first successful diagonal move is considered. However, for a H-type amino acid, the first successful diagonal move that does not increase the Cartesian distance of the amino acid from the HCC is taken. All the amino acids are traversed and the successful moves are applied as one composite move. Nevertheless, the macro-mutation squeezes the conformation and quickly forms the H-core by repeating the procedure. In our GA, macro-mutation is used like other mutation operators. Algorithm 5 presents the *pseudocode* of the macro-mutation operator.

*4.2. Stagnation Recovery.* In genetic algorithms, the similarity among individuals within the population increase as generations after generations pass on. This characteristic pushes the search towards stagnation. Any premature convergence also leads the search towards stagnation. To deal with a stagnation situation, we remove identical individuals and apply random-walk as described below.

*4.2.1. Removing Duplicates.* In our approach, unlike [36], we remove duplicates from each generation to maintain the diversity of the population. During exhaustive generation, we check the existence of the newly generated child in

```
(1) crossbredadd(conf conf′)
(2) foreach 1 ≤ pos ≤ seqLength do
(3)     c, c′ ⟵ applyOperator(conf conf′ pos)
(4)     crossbredadd(c, c′)
(5) return best2Conformations(crossbred)
```

ALGORITHM 4: The *pseudocode* of exhaustive crossover: crossConfs(conf, conf′).

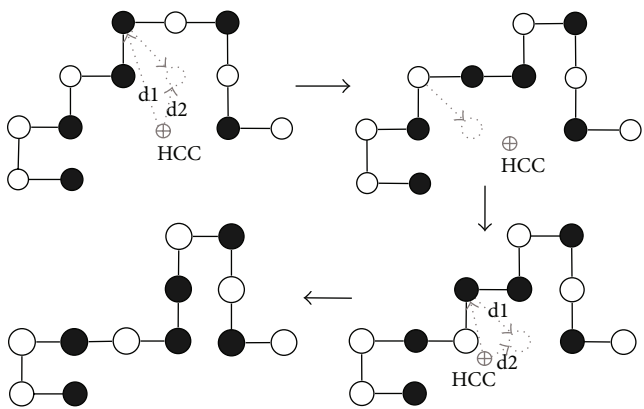

FIGURE 3: A macro-mutation operator comprising a series of diagonal moves. For simplification and easy understanding, the figures are presented in 2D space.

the new population. If it does not exist then the new solution is added to the new population list. By doing this, our approach reduces the frequency of stagnations.

*4.2.2. Applying Random Walk.* Premature H-cores are observed at search stagnation. To handle stagnation situation, in our GA (Algorithm 2 Line 18), a random-walk [46] algorithm is applied. The *pseudocode* of the algorithm is shown in Algorithm 6. This algorithm uses pull moves [47] (as shown in Figure 4(d)) to break the H-core. We use pull moves because they are complete, local, and reversible. Successful pull moves never generate infeasible conformations. During pulling, energy level and structural diversification are observed with a view to maintaining a balance between these two. We allow the energy level to change within 5% to 10% with changes in the structure from 10% to 75% of the original. We try to accept the conformation that is close to the current conformation, in terms of the energy level but as far as possible in structural diversity, and which is determined by the function checkDiversity (Algorithm 6 Line 5). In our genetic algorithm, random-walk is very effective [46] in recovering from stagnation.

*4.3. Further Implementation Details.* Our GA requires a representation of the conformations, initialisation of the population, evaluation of the solution in each iteration, and a set of genetic operators on the conformations.

*4.3.1. Representation.* We represent each conformation by using the relative encoding and the Cartesian coordinates of each amino acid. In the crossover operator, the relative encoding helps generate the offspring easily while the coordinates help in other operators particularly in finding free lattice points. The relative encoding is also used in removing duplicates within a given generation.

*4.3.2. Initialisation.* Our GA starts with an initial population, which is a set of feasible conformations. We generate initial conformations following a self-avoiding walk on FCC lattice points. The *pseudocode* of the initialisation algorithm is presented in Algorithm 7. It places the first amino acid at (0, 0, 0). It then randomly selects a basis vector to place the successive amino acid at a neighbouring free lattice point. The mapping proceeds until a self-avoiding walk is found for the whole protein sequence within a given number of iterations. If no valid conformation is found within the given number of iterations, a deterministic structure is returned. However, we have not encountered this situation in our experiments.

*4.3.3. Evaluation.* After each iteration, the conformation is evaluated by calculating the contact (topological neighbour) potentials where the two amino acids are non-consecutive. The *pseudocode* in Algorithm 8 presents the procedure of calculating the interaction energy of a given conformation. The contact potentials are found in BM energy model.

*4.3.4. Implementing Primitive Operators.* Along with exhaustiveness, macro-mutation and random-walk, the primitive operators (as shown in Figure 4) are implemented in GA. The operators we implemented are single-point crossover (Figure 4(a)), rotation mutation (Figure 4(b)), diagonal moves (Figure 4(c)), pull moves (Figure 4(d)), and tilt moves (Figure 4(e)). Rotation, diagonal moves, pull moves, and tilt moves are implemented as mutation operators.

(1) Crossover: at a given crossover point (dotted circle in Figure 4(a)), two parent conformations exchange their parts and generate two children. The success rate of the crossover operator decreases with the increase in the compactness of the structure.

(2) Rotation: one part of a given conformation is rotated around a selected point (Figure 4(b)). This move is mostly effective at the beginning of the search.

(3) Diagonal move: given three consecutive amino acids at lattice points A, B, and C, a diagonal move at position B takes the corresponding amino acid diagonally

```
(1)  return conf
(2)  for i = 1 to repeat do
(3)      T ⟵ P if bernoulli(p), else H
(4)      AA[j] : jth amino acid in conformation
(5)      point: unoccupied new position for AA[j]
(6)      hcc ⟵ findHCC ()
(7)      foreach j: type(AA[j]) = T do
(8)          d_old ⟵ getDistance (AA[j], hcc)
(9)          if T = P then
(10)             point ⟵ findFreePoint (AA[j])
(11)             applyDiagonalMove (AA[j], point)
(12)         else
(13)             point ⟵ findFreePoint (AA[j])
(14)             d_new ⟵ getDistance (point, hcc)
(15)             if d_new ≤ d_old then
(16)                 applyDiagonalMove (AA[j], point)
(17)                 break
```

ALGORITHM 5: The *pseudocode* of macro-mutation operator: macroMutation(AA, repeat).

```
(1) isFound ⟵ false
(2) while (!isFound) do
(3)     for (pos = 1 to seqLength) do
(4)         applyPullMove(pos)
(5)     isFound ⟵ checkDiversity()
```

ALGORITHM 6: The *pseudocode* of random-walk technique: randomWalk( ).

```
(1) AA[0] ⟵ AminoAcid(0,0,0)
(2) for a number of times do
(3)     for (i = 1 to seqLength – 1) do
(4)         k ⟵ getRandom(12)
(5)         node ⟵ AA[i – 1] + basisVec[k]
(6)         if node is not free then break
(7)         AA[i] ⟵ AminoAcid(node)
(8)     if full structure found then return AA[]
(9) return AA[] having a deterministic structure;
```

ALGORITHM 7: The *pseudocode* of random initialisation: initialise( ).

```
(1) fitness ⟵ 0
(2) for (i = 0 to seqLength – 1) do
(3)     for (j = i + 2 to seqLength – 1) do
(4)         nodeI ⟵ AA[i]
(5)         nodeJ ⟵ AA[j]
(6)         sqrD ⟵ getSqrDist(nodeI, nodeJ)
(7)         if sqrD = 2 then
(8)             fitness ⟵ fitness + E_bm[i][j]
(9) return fitness
```

ALGORITHM 8: The *pseudocode* of evaluation procedure: evaluate(AA).

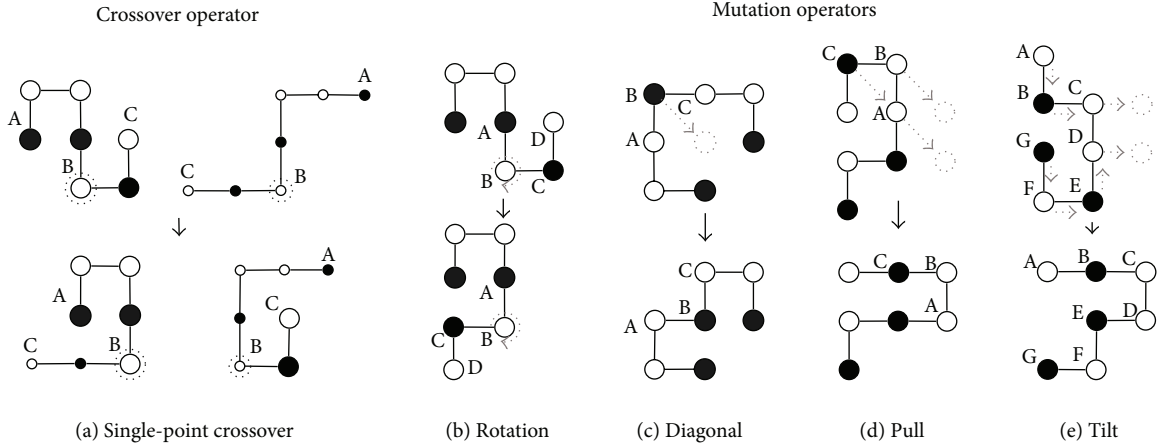

(a) Single-point crossover    (b) Rotation    (c) Diagonal    (d) Pull    (e) Tilt

FIGURE 4: The operators that are used in our GA on 3D FCC lattice space. For simplification and easy understanding, the figures are presented in 2D space. The black solid circles represent the hydrophobic amino acids and others are hydrophilic.

to a free position (Figure 4(c)). Diagonal moves are very effective on FCC lattice [32, 34] points.

(4) Pull moves: the amino acids at points A and B are pulled to the free points (Figure 4(d)) and the connected amino acids are pulled as well to get a valid conformation. Pull moves [47] are local, complete and reversible. Pull moves are very effective especially when the conformation is compact.

(5) Tilt moves: Two or more consecutive amino acids connected in a straight line are moved by a tilt move to immediately parallel lattice positions [25]. Tilt moves pull the conformation from both sides until a valid conformation is found. In Figure 4(e), the amino acids at points C and D are moved and subsequently other amino acids from both sides are moved as well.

## 5. Results and Analyses

We implemented our GA framework in Java (J2EE). We ran our experiments on the NICTA (NICTA website: http://www.nicta.com.au/) cluster. The cluster consists of a number of identical Dell PowerEdge R415 computers, each equipped with 2× AMD 6-Core Opteron 4184 processors, 2.8 GHz clock speed, 3 M L2/6 M L3 cache, and 64 GB memory and running Rocks OS (a Linux variant for cluster).

*5.1. Benchmark.* The protein instances used in our experiments are taken from the literature (as shown in Table 2). The first seven proteins *4RXN, 1ENH, 4PTI, 2IGD, 1YPA, 1R69,* and *1CTF* are taken from [12] and the next five proteins *3MX7, 3NBM, CMQO, 3MRO,* and *3PNX* from [13].

*5.2. Comparison with the State-of-the-Art Results.* Table 3 compares our experimental results with the state-of-the-art results obtained by [12, 13] on FCC lattice and BM energy model. We compare our results using two criteria: free energy level obtained by our algorithm and that obtained by other algorithms, and the root mean square deviation (RMSD) of the structures obtained by our algorithm and that obtained by other algorithms.

*5.2.1. Interaction Energy Level.* We present interaction energy values in two different formats: the global lowest interaction energy (column *Best*) and the average (column *Avg*) of the lowest interaction energies obtained from 50 different runs. In case of the global best energy, our approach outperforms the state-of-the-art approaches in [12, 13] in 9 out of 12 benchmark proteins. However, in case of average energy, our approach outperforms both of the approaches in all 12 benchmark proteins.

*5.2.2. Root Mean Square Deviation.* The root mean square deviation (RMSD) is frequently used in measuring the difference between values predicted by a model and the values actually observed. For any given structure, the RMSD is calculated using

$$\text{RMSD} = \sqrt{\frac{\sum_{i=1}^{n-1}\sum_{j=i+1}^{n}\left(d_{ij}^{p}-d_{ij}^{n}\right)^{2}}{n*(n-1)/2}}, \quad (4)$$

where $d_{ij}^{p}$ and $d_{ij}^{n}$ denote the distances between $i$th and $j$th amino acids, respectively, in the predicted structure and the native structure of the protein. To calculate the RMSD, the distance between two neighbour lattice points ($\sqrt{2}$ for FCC lattice) is considered as 3.8 Å, the average distance between two $\alpha$-Carbons. From the experimental results, it is clear that our approach significantly improves the RMSD values in comparison to the state-of-the-art approaches [12, 13].

*5.2.3. Statistical Significance Test.* For both average interaction energy and average RMSD values, we performed the Mann-Whitney $U$ test with 95% confidence level to verify the significance of difference between our results and the state-of-the-art results. The test outcomes supported our improvements in all 12 proteins.

Table 2: The benchmark proteins used in our experiments.

| ID | Length | Sequence |
|---|---|---|
| 4RXN | 54 | MKKYTCTVCGYIYNPEDGDPDNGVNPGTDFKDIPDDWVCPLCGVGKDQFEEVEE |
| 1ENH | 54 | RPRTAFSSEQLARLKREFNENRYLTERRRQQLSSELGLNEAQIKIWFQNKRAKI |
| 4PTI | 58 | RPDFCLEPPYTGPCKARIIRYFYNAKAGLCQTFVYGGCRAKRNNFKSAEDCMRTCGGA |
| 2IGD | 61 | MTPAVTTYKLVINGKTLKGETTTKAVDAETAEKAFKQYANDNGVDGVWTYDDATKTFTVTE |
| 1YPA | 64 | MKTEWPELVGKAVAAAKKVILQDKPEAQIIVLPVGTIVTMEYRIDRVRLFVDKLDNIAQVPRVG |
| 1R69 | 69 | SISSRVKSKRIQLGLNQAELAQKVGTTQQSIEQLENGKTKRPRFLPELASALGVSVDWLLNGTSDSNVR |
| 1CTF | 74 | AAEEKTEFDVILKAAGANKVAVIKAVRGATGLGLKEAKDLVESAPAALKEGVSKDDAEALKKALEEAGAEVEVK |
| 3MX7 | 90 | MTDLVAVWDVALSDGVHKIEFEHGTTSGKRVVYVDGKEEIRKEWMFKLVGKETFYVGAAKTKATINIDAISGFA YEYTLEINGKSLKKYM |
| 3NBM | 108 | SNASKELKVLVLCAGSGTSAQLANAINEGANLTEVRVIANSGAYGAHYDIMGVYDLIILAPQVRSYYREMKVDAE RLGIQIVATRGMEYIHLTKSPSKALQFVLEHYQ |
| 3MQO | 120 | PAIDYKTAFHLAPIGLVLSRDRVIEDCNDELAAIFRCARADLIGRSFEVLYPSSDEFERIGERISPVMIAHGSYADDR IMKRAGGELFWCHVTGRALDRTAPLAAGVWTFEDLSATRRVA |
| 3MRO | 142 | SNALSASEERFQLAVSGASAGLWDWNPKTGAMYLSPHFKKIMGYEDHELPDEITGHRESIHPDDRARVLAALK AHLEHRDTYDVEYRVRTRSGDFRWIQSRGQALWNSAGEPYRMVGWIMDVTDRKRDEDALRVSREELRRL |
| 3PNX | 160 | GMENKKMNLLLFSGDYDKALASLIIANAAREMEIEVTIFCAFWGLLLLRDPEKASQEDKSLYEQAFSSLTPREAE ELPLSKMNLGGIGKKMLLEMMKEEKAPKLSDLLSGARKKEVKFYACQLSVEIMGFKKEELFPEVQIMDVKEYLK NALESDLQLFI |

Table 3: The energy and RMSD values achieved by different algorithms using BM energy model. The bold-faced values indicate the winner. For both energy and RMSD values, the higher the better.

| Protein size and H-Count | | | Hybrid [12] | | | Heuristics [13] ($r$) | | | Our GA ($t$) | | | | Our Rel. Imp. RI w.r.t [13] | |
|---|---|---|---|---|---|---|---|---|---|---|---|---|---|---|
| | | | BM En | | RMSD | BM En | | RMSD | BM En | | RMSD | | | |
| Seq | Size | H | Best | Avg | Avg | Best | Avg | Avg | Best | Avg | Best | Avg | Energy | RMSD |
| 4RXN | 54 | 27 | −157.70 | −140.13 | 9.99 | −165.21 | −156.32 | 6.29 | **−166.88** | **−162.72** | 4.70 | **5.41** | 4.09% | 13.99% |
| 1ENH | 54 | 19 | −154.24 | −141.99 | 10.04 | **−168.75** | −146.69 | 6.61 | −153.79 | **−151.65** | 4.57 | **5.22** | 3.01% | 21.03% |
| 4PTI | 58 | 32 | −213.70 | −196.23 | 11.92 | **−219.52** | −198.42 | 7.07 | −210.29 | **−204.56** | 5.97 | **6.46** | 3.09% | 36.92% |
| 2IGD | 61 | 25 | −184.29 | −157.20 | 13.30 | **−187.20** | −174.19 | 9.33 | −183.18 | **−176.83** | 6.85 | **7.81** | 1.12% | 16.26% |
| 1YPA | 64 | 38 | −221.11 | −208.10 | 13.42 | −249.90 | −239.98 | 7.53 | **−256.95** | **−253.09** | 5.42 | **6.29** | 5.46% | 16.47% |
| 1R69 | 69 | 30 | −180.62 | −165.11 | 14.78 | −213.04 | −204.17 | 6.47 | **−216.37** | **−208.79** | 4.68 | **5.17** | 2.26% | 20.09% |
| 1CTF | 74 | 42 | −204.88 | −195.23 | 12.65 | −224.29 | −213.81 | 7.23 | **−233.51** | **−225.43** | 4.69 | **5.28** | 5.43% | 26.97% |
| 3MX7 | 90 | 44 | — | — | — | −328.12 | −311.56 | 8.18 | **−340.05** | **−325.45** | 7.31 | **7.94** | 4.46% | 2.93% |
| 3NBM | 108 | 56 | — | — | — | −418.60 | −401.99 | 8.58 | **−436.76** | **−419.25** | 5.58 | **6.46** | 4.29% | 24.71% |
| 3MQO | 120 | 68 | — | — | — | −465.74 | −455.27 | 8.86 | **−486.05** | **−472.78** | 6.17 | **6.84** | 3.85% | 22.80% |
| 3MRO | 142 | 63 | — | — | — | −445.33 | −430.29 | 10.02 | **−479.36** | **−447.77** | 7.65 | **8.72** | 4.06% | 12.97% |
| 3PNX | 160 | 84 | — | — | — | −601.23 | −571.13 | 9.38 | **−615.82** | **−592.25** | 7.50 | **8.51** | 3.70% | 9.28% |

*5.3. Relative Improvement (RI).* The difficulty to improve energy level is increased as the predicted energy level approaches to a known lower bound of a given protein. For example, if the lower bound of free energy of a protein is −100, the efforts to improve energy level from −80 to −85 is much less than that to improve energy level from −95 to −100 though the change in energy is the same (−5). The rightmost RI columns in Table 3 show the relative improvements that our algorithm (target) achieved with respect to the state-of-the-art approaches (reference). For each protein, the relative improvement of the target ($t$) with respect to the reference ($r$) is calculated using (5), where $E_t$ and $E_r$ denote the average energy values achieved by target and reference, respectively. We use a similar equation to calculate the relative improvement for RMSD:

$$\text{RI} = \frac{E_t - E_r}{E_r} * 100\%. \quad (5)$$

From these results, we see that our GA in all 12 proteins improves the search quality from 1.12% to 5.46% in the average interaction energy and from 2.93% to 36.92% in the RMSD values.

*5.4. Detailed Analyses.* The BM energy model actually implicitly bears the characteristic of hydrophobicity. The matrix values present some variations within amino acids of the same class (H or P). A partition algorithm such as 2-means clustering algorithm easily reveals the H-P partitioning within the BM model. Given this knowledge, we study the effect of explicitly using hydrophobic property within our GA.

*5.5. Effect of Macromutation Operator.* Our macro-mutation operator biases the search towards a hydrophobic core by applying a series of diagonal moves and thus achieves improvements in terms of BM energy values of the output conformations. One question to be investigated is whether the improvement comes only from the repeated application of diagonal moves but not from the exploitation of the hydrophobicity knowledge. If hydrophobicity is the reason for improvement, a related question is whether using solely the HP model instead of using BM model anywhere in the search will be useful; the output energy level of course will be in the BM model. In order to answer these questions, we implemented four different versions of our genetic algorithm.

(1) BH: this version is our final algorithm that we described in detail, and used in presenting our main results in Table 3 and in comparing with the state-of-the-art results. To reiterate, this version uses the BM energy model for search and energy reporting and hydrophobicity knowledge in the macro-mutation operator that repeatedly applies diagonal moves towards forming a hydrophobic core.

(2) BD: this version of our GA uses the BM energy model for search and energy reporting. It also uses the BM energy model within the macro-mutation operator that applies diagonal moves repeatedly. This version will show the effect of the macro-mutation without using the hydrophobic property.

(3) BM: this version of our GA uses the BM energy model for search and energy reporting. However, this version does not have any macro-mutation operator and thus could be seen as the baseline algorithm for the questions to be investigated.

(4) HP: this version of our GA uses the HP energy model for search. However, we report the energy values of the final conformations returned by the GA in BM energy model. Note that this version has the hydrophobic core directed macro-mutation operator. This version will show whether HP model is sufficient even when the energy of a conformation is to be in the BM model.

Table 4 presents the experimental results to show the effects of using hydrophobicity knowledge within the macro-mutation operator. The energy values are obtained by running each algorithm with a 60-minute time cutoff. The average values are calculated over 50 different runs. We also perform the Mann-Whitney $U$ test with 95% confidence level to test the significance of differences. From these results, we see that the HP version performs the worst followed by the BD version. The BD version itself performs worse than the BM version which is worse than the BH version. From all these, we conclude that using solely HP model is not sufficient and the application of repeated diagonal moves without using the hydrophobicity is not helpful.

For further analysis of the macro-mutation, in Table 5 we present the total numbers of contact and the numbers of H-H, H-P, and P-P contacts present in the output conformations of the above-mentioned four variants of our GA. The number of contacts are the average over the 50 runs for each protein.

Analysing the number of contacts, we see that the HP version has more H-H contacts, but fewer H-P, P-P, and total contacts than the other three versions. In the BM energy model the H-H contacts contribute the large energy values; however, unlike the H-P model, other types of contact in BM model also contribute energy values. In the BD and BM versions the improvement compared to the HP version comes from the increase in the number of H-P, P-P, and total contacts, although there is a decrease in the H-H contacts. The BD version performs worse than the BM version implying the repeated diagonal moves without using hydrophobicity are rather harmful. The BH version improves over BD and BM by increasing the H-H contacts and decreasing the H-P contacts; the total contacts and the P-P contacts remain almost similar. All these lead us to conclude that the explicit use of the implicit hydrophobicity knowledge within the macro-mutation is effective. The search algorithm could not utilise the HP knowledge implicitly buried within the BM model.

To demonstrate the search progress, we periodically find the best energy values obtained so far in each run. For a given period, we then calculate the average energy values obtained for that period over 50 runs. We used a 2-minute time interval. Figure 5 presents the average energy values obtained at each time interval for two different proteins: *4RXN* (Figure 5(a)) and *3PNX* (Figure 5(b)) are the smallest and the largest amongst the 12 benchmark proteins. From both of the charts, we see that the final version of our algorithm BH clearly outperforms the other three versions.

*5.6. Effect of Initialisation.* Initialising the population is an important part of population based search algorithms. We initialise the GA population with randomly generated valid structures using the procedure shown in Algorithm 7. As noted before, our GA uses a macro-mutation operator that tries to take the search towards forming a hydrophobic core. This leads us to test whether initialisation with conformations that already have optimal HP core at the center would expedite the search. For this, we use the CPSP tools that produce structures with very compact hydrophobic cores. Table 6 shows the results of these experiments; the reported values are average over 50 runs for each protein. We see that compared to our random initialisation method, initialisation with the structures produced by the CPSP-tools, although it gives a very good energy value at the beginning, it leads to worse performance at the end. This observation remains the same when the hydrophobic core-directed macro-mutation operator is used and when it is not used.

*5.7. Simplified Structure.* Figure 6 shows FCC structures of six different proteins at their lowest free energy levels as obtained by our final GA variant. We use Jmol (Jmol: an open-source Java viewer for chemical structures in 3D. http://www.jmol.org/) to draw the structures.

## 6. Conclusion

In this paper, we presented a genetic algorithm for protein structure prediction on 3D face-centred-cubic lattice. Our algorithm mainly uses a $20 \times 20$ energy matrix as its energy model but also incorporates the hydrophobic-polar model to bias the search towards exploring the structures that have hydrophobic cores. The bias is obtained by applying a hydrophobic-core directed macro-mutation operator. By using the two energy models in a mixed fashion, our algorithm significantly outperforms the state-of-the-art approaches for the similar models in terms of lower interaction energies and lower RMSD values. A concern often raised in PSP is about the usefulness of a low resolution energy model when the target is to obtain realistic structures in the high resolution model. Our results show that a low resolution energy model could be useful even when a high resolution energy model is used in the problem. In future, we intend to apply our approach in real model based protein structure prediction.

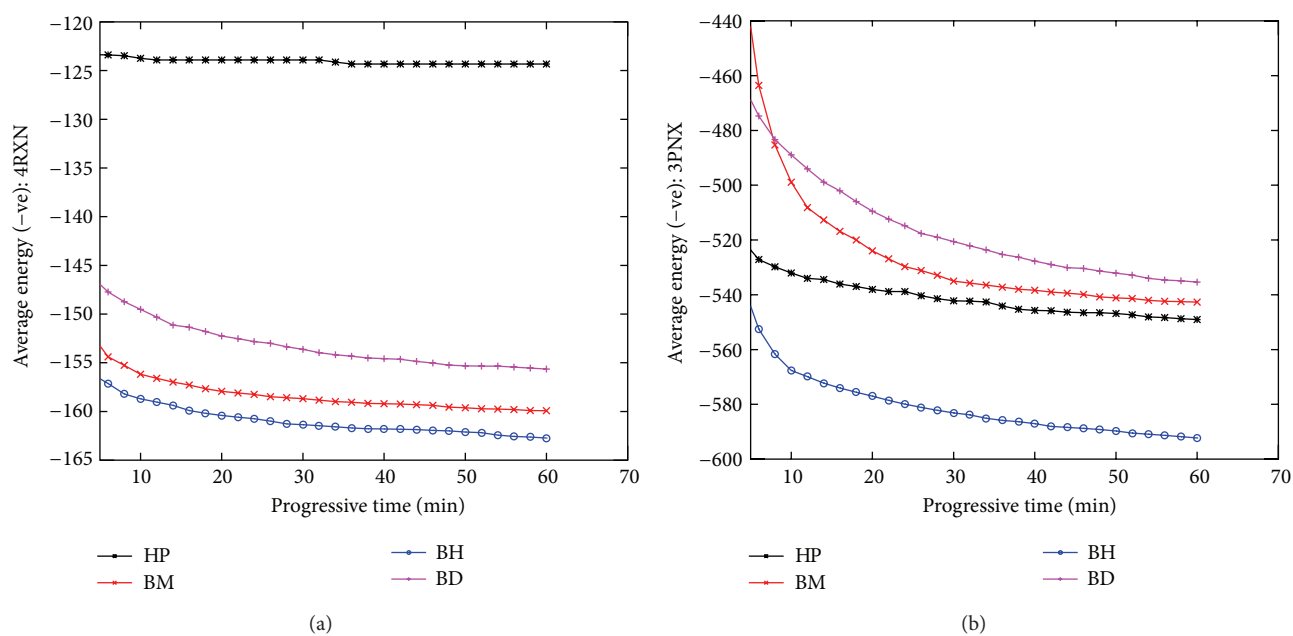

Figure 5: The search progress over time for two proteins of different size.

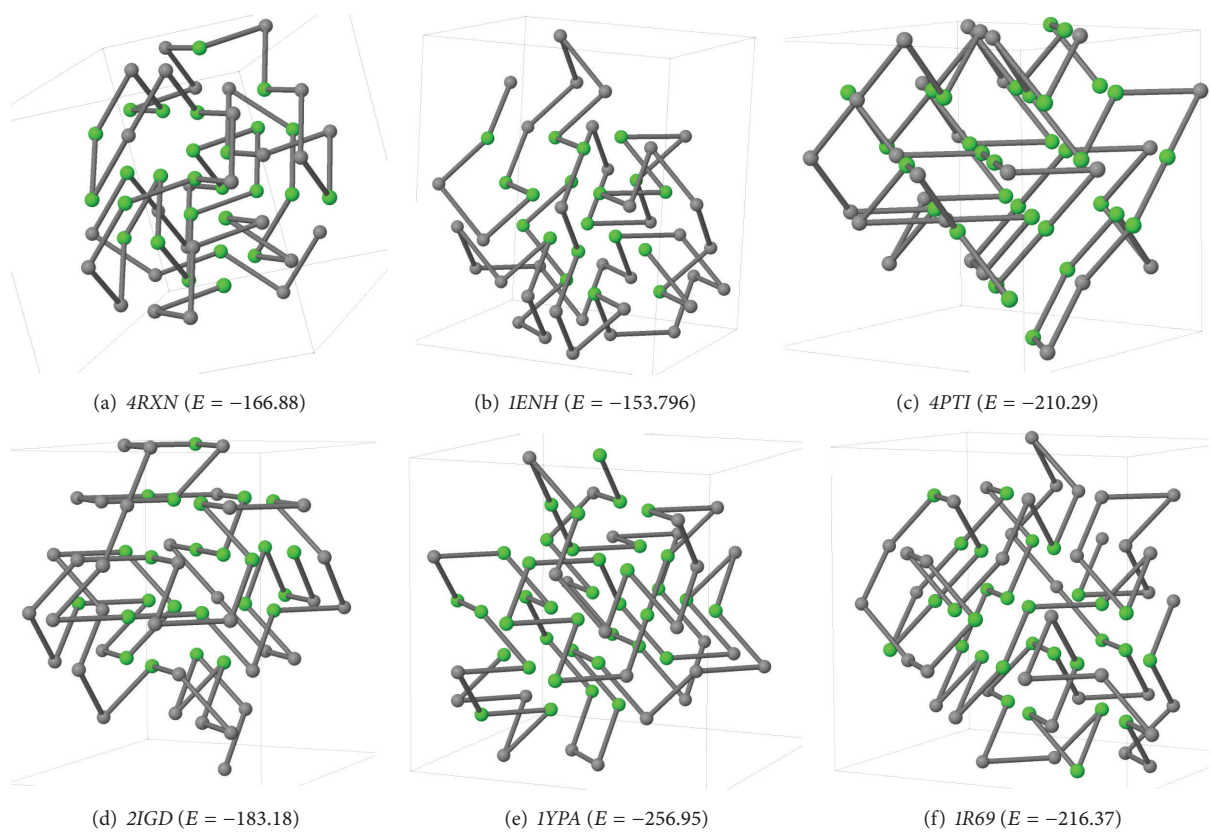

Figure 6: Simplified structure of six different proteins.

TABLE 4: The effect of using HP energy model within a macro-mutation operator. The bold-faced values indicate the winner. The lower the energy value, the better the performance.

| Protein details | | | Best of 50 runs | | | | Average of 50 runs | | | | Rel. Imp. RI | |
|---|---|---|---|---|---|---|---|---|---|---|---|---|
| Seq | Size | H | HP | BM | BD | BH | HP($r$) | BM($r$) | BD | BH($t$) | HP | BM |
| 4RXN | 54 | 27 | −135.43 | **−167.90** | −162.81 | −166.88 | −124.32 | −159.92 | −155.64 | **−162.72** | 30.89% | 1.75% |
| 1ENH | 54 | 19 | −134.97 | **−155.49** | −150.07 | −153.79 | −125.52 | −147.42 | −144.47 | **−151.65** | 20.82% | 2.87% |
| 4PTI | 58 | 32 | −171.28 | **−213.05** | −202.33 | −210.29 | −156.43 | −201.12 | −196.86 | **−204.56** | 30.77% | 1.71% |
| 2IGD | 61 | 25 | −153.00 | −181.93 | −177.19 | **−183.18** | −140.59 | −173.62 | −170.79 | **−176.83** | 25.78% | 1.85% |
| 1YPA | 64 | 38 | −232.94 | −255.40 | −251.78 | **−256.95** | −220.35 | −247.17 | −242.89 | **−253.09** | 14.86% | 2.4% |
| 1R69 | 69 | 30 | −181.44 | −212.35 | −213.34 | **−216.37** | −171.79 | −203.26 | −199.65 | **−208.79** | 21.54% | 2.72% |
| 1CTF | 74 | 42 | −202.06 | −225.59 | −225.37 | **−233.51** | −190.31 | −217.02 | −212.05 | **−225.42** | 18.45% | 3.87% |
| 3MX7 | 90 | 44 | −295.16 | −333.74 | −323.67 | **−340.05** | −281.99 | −317.11 | −311.92 | **−325.45** | 15.41% | 2.63% |
| 3NBM | 108 | 56 | −380.20 | −426.35 | −424.10 | **−436.76** | −364.99 | −406.11 | −400.17 | **−419.25** | 14.87% | 3.24% |
| 3MQO | 120 | 68 | −443.84 | −472.15 | −464.09 | **−486.05** | −420.38 | −452.32 | −443.08 | **−472.78** | 12.46% | 4.52% |
| 3MRO | 142 | 63 | −420.65 | −445.19 | −444.99 | **−479.36** | −401.32 | −420.86 | −421.61 | **−447.77** | 11.57% | 6.39% |
| 3PNX | 160 | 84 | −576.77 | −584.17 | −576.09 | **−615.82** | −549.03 | −542.68 | −535.40 | **−592.25** | 7.87% | 9.13% |

TABLE 5: The number of H-H, H-P, and P-P contacts present in the output conformations.

| Protein details | | | H-H contacts | | | | H-P contacts | | | | P-P contacts | | | | Total contacts | | | |
|---|---|---|---|---|---|---|---|---|---|---|---|---|---|---|---|---|---|---|
| Seq | Size | H | HP | BM | BD | BH | HP | BM | BD | BH | HP | BM | BD | BH | HP | BM | BD | BH |
| 4RXN | 54 | 27 | 76 | 62 | 58 | 67 | 36 | 57 | 56 | 54 | 17 | 22 | 24 | 22 | 131 | 142 | 139 | 143 |
| 1ENH | 54 | 19 | 51 | 44 | 40 | 47 | 37 | 59 | 57 | 56 | 35 | 42 | 44 | 40 | 124 | 145 | 142 | 145 |
| 4PTI | 58 | 32 | 92 | 74 | 69 | 77 | 40 | 64 | 65 | 63 | 15 | 21 | 22 | 20 | 148 | 160 | 157 | 161 |
| 2IGD | 61 | 25 | 71 | 54 | 49 | 62 | 46 | 72 | 71 | 66 | 33 | 40 | 41 | 40 | 151 | 167 | 162 | 168 |
| 1YPA | 64 | 38 | 117 | 104 | 101 | 109 | 38 | 55 | 55 | 52 | 15 | 17 | 18 | 17 | 171 | 178 | 175 | 179 |
| 1R69 | 69 | 30 | 90 | 73 | 69 | 80 | 49 | 77 | 74 | 72 | 38 | 43 | 45 | 44 | 177 | 194 | 190 | 197 |
| 1CTF | 74 | 42 | 131 | 115 | 110 | 122 | 46 | 64 | 64 | 60 | 20 | 22 | 24 | 22 | 198 | 203 | 199 | 205 |
| 3MX7 | 90 | 44 | 140 | 109 | 103 | 121 | 68 | 106 | 106 | 98 | 44 | 45 | 48 | 46 | 254 | 262 | 258 | 266 |
| 3NBM | 108 | 56 | 183 | 137 | 132 | 153 | 79 | 140 | 137 | 131 | 49 | 53 | 56 | 53 | 312 | 331 | 327 | 338 |
| 3MQO | 120 | 68 | 227 | 180 | 169 | 201 | 88 | 139 | 144 | 128 | 45 | 52 | 55 | 53 | 361 | 372 | 369 | 383 |
| 3MRO | 142 | 63 | 206 | 143 | 134 | 172 | 113 | 185 | 181 | 172 | 98 | 106 | 115 | 110 | 418 | 435 | 431 | 455 |
| 3PNX | 160 | 84 | 280 | 219 | 202 | 253 | 137 | 183 | 176 | 176 | 80 | 70 | 73 | 77 | 499 | 472 | 452 | 507 |

TABLE 6: Effect of initialisation.

| Protein details | | | Initialised by CPSP tools | | | | Initialised by Algorithm 7 | | | |
|---|---|---|---|---|---|---|---|---|---|---|
| | | | BM | | BH | | BM | | BH | |
| Seq | Size | H | Start | End | Start | End | Start | End | Start | End |
| 4RXN | 54 | 27 | −127.80 | −152.87 | −127.80 | −156.50 | −77.13 | −159.92 | −60.84 | −162.72 |
| 1ENH | 54 | 19 | −130.13 | −142.18 | −130.13 | −149.93 | −67.37 | −147.42 | −54.34 | −151.65 |
| 4PTI | 58 | 32 | −156.19 | −193.42 | −156.19 | −198.80 | −95.99 | −201.12 | −79.42 | −204.56 |
| 2IGD | 61 | 25 | −139.00 | −170.17 | −139.00 | −171.35 | −80.38 | −173.62 | −65.81 | −176.83 |
| 1YPA | 64 | 38 | −226.67 | −241.17 | −226.67 | −247.62 | −117.15 | −247.17 | −97.94 | −253.09 |
| 1CTF | 74 | 42 | −189.68 | −213.44 | −189.68 | −222.96 | −89.86 | −217.02 | −72.20 | −225.42 |

## Conflict of Interests

The authors declare that they have no competing interests.

## Authors' Contribution

Mahmood A. Rashid conceived the idea of applying *mixed energy model* in genetic algorithms. M. A. Hakim Newton, Md. Tamjidul Hoque, and Abdul Sattar helped M. A. Rashid modeling, implementing, and testing their algorithm. All authors equally participated in analysing the test results to improve the algorithm and were significantly involved in the process of writing and reviewing the paper.

## Acknowledgments

The authors would like to express their great appreciation to the people managing the *Cluster Computing Services* at National ICT Australia (NICTA) and Griffith university. Md. Tamjidul Hoque acknowledges the Louisiana Board of Regents through the Board of Regents Support Fund, *LEQSF (2013-16)-RD-A-19*. NICTA, the sponsor of the paper for publication, is funded by the Australian Government as represented by the Department of Broadband, Communications and the Digital Economy and the Australian Research Council through the ICT Centre of Excellence program.

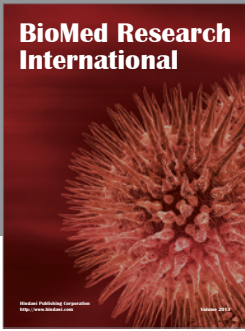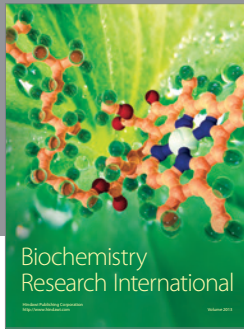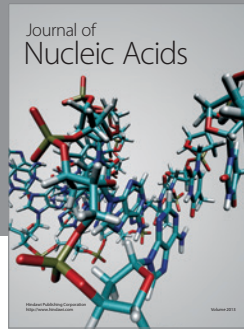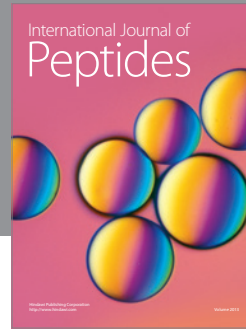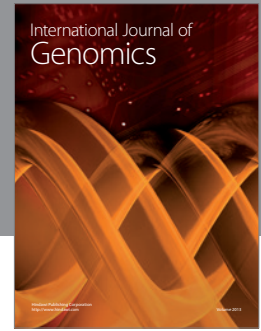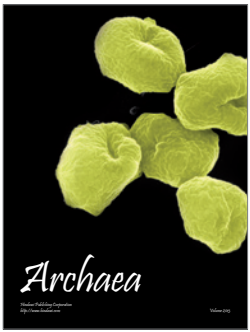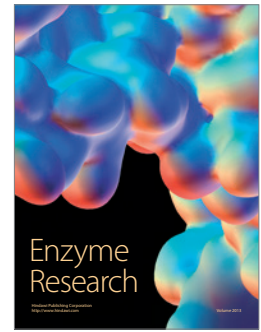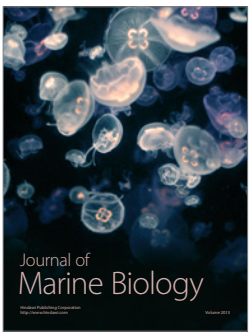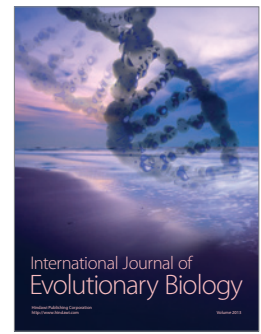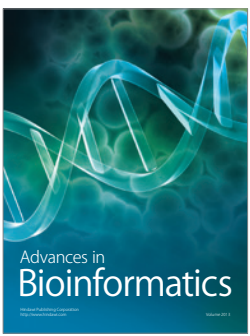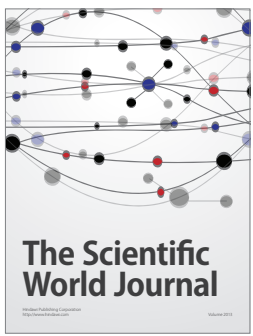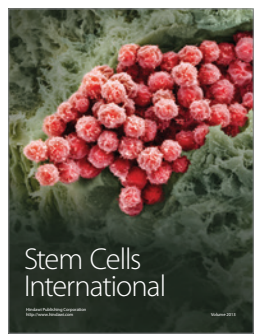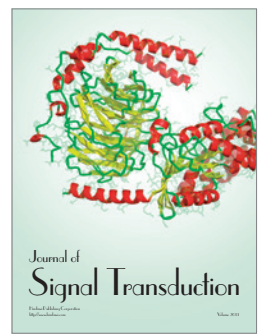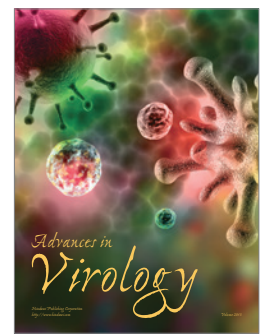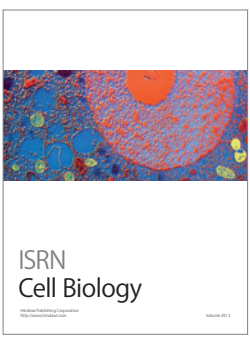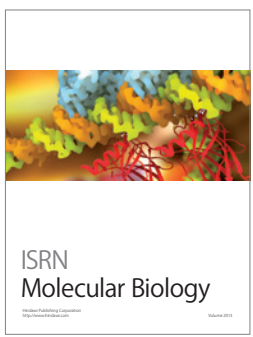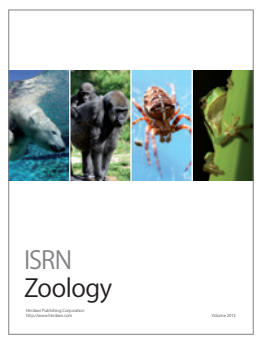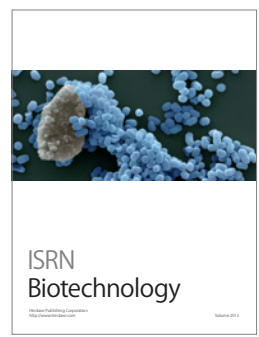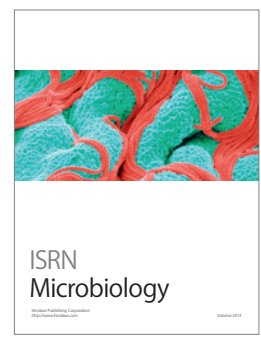